\documentclass[twocolumn,aps,pra,showpacs,tightenlines]{revtex4-1}
\usepackage{amsmath}
\usepackage{amsfonts}
\usepackage{graphicx}
\usepackage{epsfig}
\usepackage{color}
\usepackage[colorlinks,citecolor=blue]{hyperref}

\begin{document}

\title{Spectral characterization of couplings in a mixed optomechanical model}
\author{Yue-Hui Zhou}
\affiliation{Key Laboratory of Low-Dimensional Quantum Structures and Quantum Control of Ministry of Education, Department of Physics and Synergetic Innovation Center for Quantum Effects and Applications, Hunan Normal University, Changsha 410081, China}
\author{Fen Zou}
\affiliation{Key Laboratory of Low-Dimensional Quantum Structures and Quantum Control of Ministry of Education, Department of Physics and Synergetic Innovation Center for Quantum Effects and Applications, Hunan Normal University, Changsha 410081, China}
\author{Xi-Ming Fang}
\affiliation{Key Laboratory of Low-Dimensional Quantum Structures and Quantum Control of Ministry of Education, Department of Physics and Synergetic Innovation Center for Quantum Effects and Applications, Hunan Normal University, Changsha 410081, China}
\author{Jin-Feng Huang}
\email{jfhuang@hunnu.edu.cn}
\affiliation{Key Laboratory of Low-Dimensional Quantum Structures and Quantum Control of Ministry of Education, Department of Physics and Synergetic Innovation Center for Quantum Effects and Applications, Hunan Normal University, Changsha 410081, China}
\author{Jie-Qiao Liao}
\email{jqliao@hunnu.edu.cn}
\affiliation{Key Laboratory of Low-Dimensional Quantum Structures and Quantum Control of Ministry of Education, Department of Physics and Synergetic Innovation Center for Quantum Effects and Applications, Hunan Normal University, Changsha 410081, China}

\begin{abstract}
We study the spectrum of single-photon emission and scattering in a mixed optomechanical model which consists of both linear and quadratic optomechanical interactions. The spectra are calculated based on the exact long-time solutions of the single-photon emission and scattering processes in this system. We find that there exist some phonon sideband peaks in the spectra and there are some sub peaks around the phonon sideband peaks under proper parameter conditions. The correspondence between the spectral features and the optomechanical interactions is confirmed, and the optomechanical coupling strengths can be inferred by analyzing the resonance peaks and dips in the spectra.
\end{abstract}


\date{\today}
\maketitle

\section{Introduction}
The optomechanical interactions between the photons and the mechanical oscillation are at the heart of the field of cavity optomechanics~\cite{Kippenberg2008Science,Aspelmeyer2012PhysTod,Aspelmeyer2014RMP,Bowen2016book}. Typically, there are two kinds of optomechanical interactions: the linear (i.e., the radiation-pressure-type) optomechanical coupling~\cite{Law1995PRA,Bose1997PRA,Vitali2007PRL,Wilson-Rae2007PRL,Marquardt2007PRL,Teufel2011Nature,Chan2011Nature,Tian2013PRL,Wang2013PRL} and the quadratic optomechanical coupling~\cite{Harris2008Nature,Meystre2008PRA,Meystre2008PRAB,Agarwal2008PRA,Jayich2008NJP,Harris2010Nature,Liao2014SC,Liao2013PRA}. In the two cases, the optomechanical couplings depend linearly and quadratically on the mechanical displacement, respectively. So far, much effort has been devoted to the studies of linear and quadratic optomechanical effects and the applications of optomechanical interactions to modern quantum technologies~\cite{Metcalfe2014APR}, including the demonstration of the fundamental of quantum theory~\cite{Aspelmeyer2012PhysTod} and the applications of optomechanical systems in quantum precision measurement~\cite{Metcalfe2014APR}.

Motivated by the great advances in the enhancement of the optomechanical coupling at the level of single photons~\cite{Gupta2007PRL,Brennecke2008Science,Eichenfield2009Nature,Pirkkalaine2015NatComm}, much recent interest has been paid to the study of the single-photon strong-coupling regime of cavity optomechanics~\cite{Rabl2011PRL,Nunnenkamp2011PRL,Liao2012PRA,Qian2012PRL,Hong2013PRA,Liao2013PRAa,Tang2014PRA,Ludwig2012PRL,Stannigel2012PRL,Kronwald2013PRA,Lu2015PRL}. The optomechanical interaction in this regime can provide a platform to study the observable optomechanical effects at the level of single photons. In particular, fruitful achievements have been obtained in the studies of various effects in linear optomechanical coupling at the few- and even single-photon levels~\cite{Rabl2011PRL,Nunnenkamp2011PRL,Liao2012PRA,Liao2013PRAa,Marshall2003PRL,Liao2016PRL}. For example, the oscillating boundary of the cavity will induce a Kerr-type optical nonlinearity and this nonlinearity has been exploited to realize photon blockade effect in the linear optomechanical cavity~\cite{Rabl2011PRL,Liao2013PRAa}. The emission spectrum of the optomechanical cavity has been studied in both the continuous-wave and wavepacket driving cases~\cite{Nunnenkamp2011PRL,Liao2012PRA}. It has been found that the phonon sideband peaks appear in the spectrum when the system works in both the single-photon strong-coupling and resolved-sideband regimes. The conditional displacement dynamics of a single photon in optomechanics has been suggested to generate quantum superposition of distinct mechanical states~\cite{Marshall2003PRL,Liao2016PRL}. In addition, some applications of the optomechanical interactions at the single-photon level have been proposed. These applications include the spectrometric reconstruction of the mechanical states~\cite{Liao2014PRA} and the spectrometric detection of weak classical forces~\cite{Zhou2019arXiv}.

Recently, much attention has been paid to the studies on the mixed optomechanical model with both linear and quadratic optomechanical couplings~\cite{Xuereb2013PRA,Zhang2014PRA,Hauer2018PRA,Brunelli2018PRA,Zhang2018PRA,Rocheleau2010Nature}. In this model, the linear and quadratic optomechanical couplings provide the physical mechanisms for conditional displacement and squeezing of the mechanical resonator, respectively. These mechanisms will create interesting physical effects such as the squeezing and cooling the mechanical mode~\cite{Xuereb2013PRA}, the harmonic generation of self-sustained oscillations~\cite{Zhang2014PRA}, the phonon quantum nondemolition measurements~\cite{Hauer2018PRA}, the unconditional preparation of nonclassical states~\cite{Brunelli2018PRA}, and the optomechanically induced transparency~\cite{Zhang2018PRA}. Such a system has also been experimentally demonstrated where the mechanical resonator is prepared and detected near its ground-state motion~\cite{Rocheleau2010Nature}.

In this paper, we study the single-photon emission and scattering spectra in the mixed optomechanical model. Within the Wigner-Weisskopf framework, we obtain the exact solution of the single photon dynamics in the emission and scattering cases where a single photon is initially inside the cavity and in a Lorentizian wavepacket outside the cavity, respectively. Based on the analytical solutions, the single-photon emission and scattering spectra are obtained. We find that there exist phonon sideband peaks in the spectrum, and around the phonon sideband main peaks, there are some sub peaks, which are caused by the frequency difference of the single-photon squeezed mechanical resonator and the free mechanical resonator. By analyzing the spectra, we characterize the spectra of the system in the single-photon strong-coupling regime by indicating the relation between the spectral features and the strengths of the linear and quadratic optomechanical interactions. As a result, the optomechanical coupling strengths can be inferred by analyzing the spectral pattern. Notice that several methods have been proposed to infer the coupling strength in optomechanical systems~\cite{Liao2012PRA,Bernad2018PRA,Wang2018PRA,Tan2018arXiv}.

The rest of this paper is organized as follows. In Sec.~\ref{model}, we introduce the mixed optomechanical model and present the Hamiltonians. In Sec.~\ref{EOM}, we consider the single-photon dynamics of the system and derive the equations of motion for these probability amplitudes in the single-photon subspace. In Secs.~\ref{SPemission} and~\ref{SPscattering}, we calculate the single-photon spectrum and discuss the spectrometric estimation of the optomechanical coupling strengths in the emission and scattering cases, respectively. A conclusion is presented in Sec.~\ref{Conclusion}.

\section{Model\label{model}}
\begin{figure}[tbp]
\center
\includegraphics[bb=45 221 393 712, width=0.47 \textwidth]{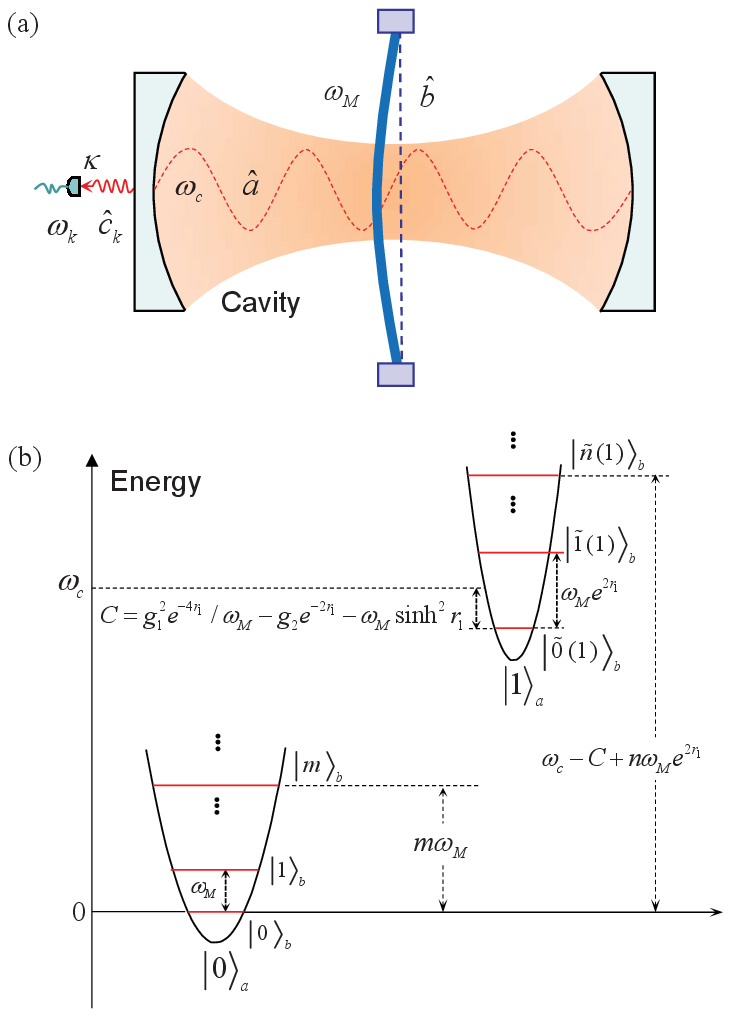}
\caption{(a) Schematic of the mixed optomechanical cavity consisting of both the linear and quadratic optomechanical interactions. The bath of the cavity is modelled as continuous fields outside the cavity. (b) The eigensystem of this mixed optomechanical cavity limited within the zero- and one-photon subspaces.}
\label{setup}
\end{figure}

We consider a mixed optomechanical model [Fig.~\ref{setup}(a)], which consists of both the linear and quadratic optomechanical interactions between the cavity field and the mechanical mode. This mixed optomechanical model is described by the Hamiltonian (with $\hbar=1$)
\begin{equation}
\hat{H}_{\text{mop}}=\omega_{c}\hat{a}^{\dagger}\hat{a}+\omega_{M}\hat{b}^{\dagger}\hat{b}
+g_{1}\hat{a}^{\dagger}\hat{a}(\hat{b}^{\dagger}+\hat{b})+g_{2}\hat{a}^{\dagger}\hat{a}(\hat{b}^{\dagger}+\hat{b})^{2},
\end{equation}
where $\hat{a}$ ($\hat{a}^{\dagger}$) and $\hat{b}$ ($\hat{b}^{\dagger}$) are, respectively, the annihilation (creation) operators of the cavity field and the mechanical mode, with the corresponding resonance frequencies $\omega_{c}$ and $\omega_{M}$. The parameters $g_{1}$ and $g_{2}$ are the single-photon coupling strengths associated with the linear and quadratic optomechanical interactions between the single-mode cavity field and the mechanical oscillation, respectively.

Physically, the single-photon emission and scattering spectra of the mixed optomechanical cavity are determined by the eigen-energy spectrum of the system within the zero- and one-photon subspaces [see Fig.~\ref{setup}(b)]. This eigen-energy spectrum can be calculated by introducing the squeezing operator $\hat{S}(\hat{r})=\exp[(\hat{r}\hat{b}^{2}-\hat{r}\hat{b}^{\dagger 2})/2]$ and the displacement operator $\hat{D}(\hat{\alpha})=\exp[\hat{\alpha}(\hat{b}^{\dagger}-\hat{b})]$, where the photon-number dependent squeezing and displacement quantities are given by
\begin{eqnarray}
\hat{r}=\frac{1}{4}\ln\left(\frac{4g_{2}}{\omega_{M}}\hat{a}^{\dagger}\hat{a}+1\right),\hspace{0.5 cm}
\hat{\alpha}=-\frac{g_{1}e^{-3\hat{r}}}{\omega_{M}}\hat{a}^{\dagger}\hat{a}.
\end{eqnarray}
Under the transformation $\hat{\tilde{H}}_{\text{mop}}=\hat{D}^{\dagger}(\hat{\alpha})\hat{S}^{\dagger}(\hat{r})\hat{H}_{\text{mop}}
\hat{S}(\hat{r})\hat{D}(\hat{\alpha})$, the Hamiltonian $\hat{H}_{\text{mop}}$ is transformed into the following diagonalized form
\begin{eqnarray}
\hat{\tilde{H}}_{\text{mop}}&=&\left(\omega_{c}+g_{2}e^{-2\hat{r}}\right)\hat{a}^{\dagger}\hat{a}+\omega_{M}\sinh ^{2}\hat{r}+\omega_{M}e^{2\hat{r}}\hat{b}^{\dagger}\hat{b}\nonumber\\
&&-\frac{g_{1}^{2}e^{-4\hat{r}}}{\omega_{M}}\hat{a}^{\dagger}\hat{a}\hat{a}^{\dagger}\hat{a}.
\end{eqnarray}

Since the photon number operator $\hat{a}^{\dagger}\hat{a}$ is a conserved quantity in $\hat{H}_{\text{mop}}$, then the squeezing and displacement quantities can be expressed as $\hat{r}=\sum_{n=0}^{\infty}r_{n}|n\rangle_{aa}\langle n|$ and $\hat{\alpha}=\sum_{n=0}^{\infty}\alpha_{n}|n\rangle_{aa}\langle n|$, with the photon-number dependent squeezing and displacement parameters $r_{n}=[\ln(4g_{2}n/\omega_{M}+1)]/4$ and $\alpha_{n}=-g_{1}e^{-3r_{n}}n/\omega_{M}$.

Denote $|n\rangle_{a}$ and $|m\rangle_{b}$ ($n,m=0,1,2,\cdots$) as the number states of the cavity field and the mechanical resonator, respectively, the eigen-system of the Hamiltonian $\hat{\tilde{H}}_{\text{mop}}$ can be obtained as
\begin{eqnarray}
\hat{\tilde{H}}_{\text{mop}}|n\rangle_{a}|m\rangle_{b}=E_{n,m}|n\rangle_{a}|m\rangle_{b},
\end{eqnarray}
where the eigenvalues are defined by
\begin{eqnarray}
E_{n,m}&=&\left(\omega_{c}+g_{2}e^{-2r_{n}}\right)n+\omega_{M}\sinh^{2}r_{n}+\omega_{M}e^{2r_{n}}m\nonumber\\
&&-\frac{g_{1}^{2}e^{-4r_{n}}}{\omega_{M}}n^{2}.
\end{eqnarray}
Then the eigen-equation of the Hamiltonian $\hat{H}_{\text{mop}}$ can be obtained as
\begin{equation}
\hat{H}_{\text{mop}}|n\rangle_{a}|\tilde{m}(n)\rangle_{b}=E_{n,m}|n\rangle_{a}|\tilde{m}(n)\rangle_{b},
\end{equation}
where
\begin{equation}
|\tilde{m}(n)\rangle_{b}=\hat{S}(r_{n})\hat{D}(\alpha_{n})|m\rangle_{b}
\end{equation}
are the squeezed displaced number states. In particular, when there is no photon in the cavity, we have $|\tilde{m}(0)\rangle_{b}=|m\rangle_{b}$.

In typical optomechanical systems, the decay rate $\gamma$ of the mechanical mode is much smaller than the decay rate $\kappa$ of the cavity field. Then in the single-photon emission and scattering processes, the mechanical damping is negligible during the time interval $1/\kappa\ll t\ll 1/\gamma$. Consequently, in the following calculations we merely consider the dissipation of the cavity field by modelling the environment of cavity field as a continuous vacuum field bath, which is described by the Hamiltonian
\begin{equation}
\hat{H}_{\text{bath}}=\sum_{k}\omega_{k}\hat{c}_{k}^{\dagger}\hat{c}_{k},
\end{equation}
where $\hat{c}_{k}$ and $\hat{c}_{k}^{\dagger}$ are the annihilation and creation operators of the $k$th mode in the bath, with the resonance frequency $\omega_{k}$. The commutation relation is given by $[\hat{c}_{k},\hat{c}_{k'}^{\dagger}]=\delta_{k,k'}$.
The coupling between the optomechanical cavity and its environment is described by the photon-hopping interaction, which takes the form as
\begin{equation}
\hat{H}_{\text{int}}=\sum_{k}\xi_{k}(\hat{a}^{\dagger}\hat{c}_{k}+\hat{c}_{k}^{\dagger}\hat{a}),
\end{equation}
with $\xi_{k}$ being the coupling strengths. Then the total Hamiltonian can be written as
\begin{equation}
\hat{H}_{\text{tot}}=\hat{H}_{\text{mop}}+\hat{H}_{\text{bath}}+\hat{H}_{\text{int}}.
\end{equation}
In the following sections, we will solve the dynamics of the total system with the probability amplitude method based on the total Hamiltonian.

\section{Equations of motion \label{EOM}}

In this system, the total photon number operator can be defined as $\hat{N}=\hat{a}^{\dagger}\hat{a}+\sum_{k}\hat{c}_{k}^{\dagger }\hat{c}_{k}$, which is a conserved quantity because of the commutative relation $[\hat{N},\hat{H}_{\text{tot}}]=0$. For studying the single-photon physics, we restrict the system within the single-photon subspace spanned over the basis states $|1\rangle_{a}|\emptyset\rangle_{c}$ and $|0\rangle_{a}|1_{k}\rangle_{c}$, which represent that a single photon inside the cavity and in the $k$th mode of the continuous-mode bath, respectively. Here, $|\emptyset\rangle_{c}$ denotes the multimode vacuum state of the continuous fields outside the cavity. A general pure state of the total system in the single-photon subspace can be expressed as
\begin{eqnarray}
|\varphi(t)\rangle&=&\sum_{n=0}^{\infty}A_{n}(t)|1\rangle_{a}|\tilde{n}(1)\rangle_{b}|\emptyset\rangle_{c}\nonumber\\
&&+\sum_{n=0}^{\infty}\sum_{k}B_{n,k}(t)|0\rangle_{a}|n\rangle_{b}|1_{k}\rangle_{c},
\end{eqnarray}
where $A_{n}(t)$ and $B_{n,k}(t)$ are the probability amplitudes.

For convenience, we will work in the rotating frame with respect to $\omega_{c}\hat{a}^{\dagger}\hat{a}$.
The Hamiltonian of the mixed optomechanical cavity in the rotating frame becomes $\hat{H}_{\text{mop}}^{I}=\omega_{M}\hat{b}^{\dagger}\hat{b}+g_{1}\hat{a}^{\dagger}\hat{a}(\hat{b}^{\dagger}+\hat{b})+g_{2}\hat{a}^{\dagger }\hat{a}(\hat{b}^{\dagger}+\hat{b})^{2}$, the eigensystem of the Hamiltonian $\hat{H}_{\text{mop}}^{I}$ is defined by
$\hat{H}_{\text{mop}}^{I}|n\rangle_{a}|\tilde{m}(n)\rangle_{b}=E^{\prime}_{n,m}|n\rangle_{a}|\tilde{m}(n)\rangle_{b}$, where the eigenvalues are given by $E^{\prime}_{n,m}=E_{n,m}-\omega_{c}n$. The total Hamiltonian in the rotating frame with respect to $\hat{H}_{0}=\omega_{c}\hat{a}^{\dagger}\hat{a}+\sum_{k}\omega_{c}\hat{c}_{k}^{\dagger}\hat{c}_{k}$ becomes
\begin{eqnarray}
\hat{H}_{I}&=&\omega_{M}\hat{b}^{\dagger}\hat{b}+g_{1}\hat{a}^{\dagger}\hat{a}
(\hat{b}^{\dagger}+\hat{b})+g_{2}\hat{a}^{\dagger }\hat{a}(\hat{b}^{\dagger}+\hat{b})^{2}\nonumber\\
&&+\sum_{k}\Delta_{k}\hat{c}_{k}^{\dagger}\hat{c}_{k}+\sum_{k}\xi_{k}
(\hat{a}^{\dagger}\hat{c}_{k}+\hat{c}_{k}^{\dagger}\hat{a}),
\end{eqnarray}
where the detuning $\Delta_{k}=\omega_{k}-\omega_{c}$ is introduced.

According to the general state $|\varphi(t)\rangle$, the Hamiltonian $\hat{H}_{I}$, and the Schr\"{o}dinger equation
$i\partial|\varphi(t)\rangle/\partial t=H_{I}|\varphi(t)\rangle$,
we obtain the equations of motion for the probability amplitudes as
\begin{subequations}
\label{eqofmprobamp}
\begin{align}
\dot{A}_{m}=&-iE_{1,m}^{\prime}A_{m}-i\sum_{n=0}^{\infty}\sum_{k}\xi_{k}\,_{b}\langle\tilde{m}(1)|n\rangle_{b}B_{n,k},\\
\dot{B}_{m,k}=&-i(E_{0,m}^{\prime}+\Delta_{k})B_{m,k}-i\sum_{n=0}^{\infty}\xi_{k}\,_{b}\langle m|\tilde{n}(1)\rangle_{b}A_{n}.
\end{align}
\end{subequations}
In the next two sections, we will solve these equations of motion for the probability amplitudes under the initial conditions corresponding to the single-photon emission and scattering cases, respectively.

\section{Single-photon emission\label{SPemission}}

In the single-photon emission case, the single photon is initially inside the cavity. With the evolution of the system, the single photon will leak out of the cavity and the emitted photon will carry the state and parameter information of the system, which can be read out from the single-photon emission spectrum. For the mechanical mode, its state could be an arbitrary state. Below, we will first solve the equations of motion~(\ref{eqofmprobamp}) corresponding to an initial number state $|m_{0}\rangle_{b}$ of the mechanical mode. In principle, the solution for general initial states of the mechanical mode can be obtained by superposition based on the solution relating to the initial number state. Note that the experimental generation of the Fock states in mechanical resonators has been reported in several systems~\cite{Connell2010Nature,Hong2017Science,Satzinger2018Nature,Chu2018Nature}. In particular, the generation of the phononic number states in cavity optomechanical systems have been theoretically proposed~\cite{Galland2014PRL,Tan2014PRA,Tan2019PRA}. For the initial state $|\varphi(0)\rangle=|1\rangle_{a}|m_{0}\rangle_{b}|\emptyset\rangle_{c}$,
the initial condition of the probability amplitudes is given by $A_{m}(0)=\,_{b}\langle\tilde{m}(1)|m_{0}\rangle_{b}$ and $B_{m,k}(0)=0$. Within the Wigner-Weisskopf framework and taking $\rho(\omega_{k})\xi_{k}^{2}=\rho(\omega_{c})\xi_{c}^{2}=\kappa/(2\pi)$ as a constant with $\rho(\omega_{k})$ being the density of the mode for the bath, the transient-state solution of these probability amplitudes can be obtained. In the long-time limit ($1/\kappa \ll t\ll 1/\gamma$), the probability amplitudes become $A_{m_{0},m}=0$ and (up to a phase factor $e^{-i(E_{0,m}^{\prime}+\Delta_{k})t})$
\begin{eqnarray}
B_{m_{0},m,k}=\sum_{n=0}^{\infty}\frac{\xi_{k}\,_{b}\langle m|\tilde{n}(1)\rangle_{b}\,_{b}\langle \tilde{n}(1)|m_{0}\rangle _{b}}{\Delta_{k}+E_{0,m}^{\prime}-E_{1,n}^{\prime}+i\kappa/2},\label{Bm0mklongt}
\end{eqnarray}
where the decay rate of the cavity field is defined by $\kappa=2\pi\rho(\omega_{c})\xi_{c}^{2}$. Notice that here we added the subscript $m_{0}$ in $A_{m_{0},m}$ and
$B_{m_{0},m,k}$ to mark the initial state $|m_{0}\rangle_{b}$ of the mechanical mode.

Accompanied with the emission of a single photon, the mixed optomechanical system transits from states $|1\rangle_{a}|\tilde{n}(1)\rangle_{b}$ to states $|0\rangle_{a}|m\rangle_{b}$. The frequency of the emitted photon is governed by the resonance condition
\begin{equation}
\Delta_{k}=E_{1,n}^{\prime}-E_{0,m}^{\prime},\label{resoncondi}
\end{equation}
which is consistent with the real part of the pole in Eq.~(\ref{Bm0mklongt}), i.e., $\Delta_{k}-(E_{1,n}^{\prime}-E_{0,m}^{\prime})=0$. The amplitude for the process
is proportional to the overlap $_{b}\langle m|\tilde{n}(1)\rangle_{bb}\langle\tilde{n}(1)\vert m_{0}\rangle_{b}$, which can be calculated with the relations
\begin{eqnarray}
_{b}\langle m|\tilde{n}(1)\rangle_{b}=\;_{b}\langle m|\hat{S}(r_{1})\hat{D}(\alpha_{1})|n\rangle_{b},\label{innerprodxx}
\end{eqnarray}
where the single-photon squeezing and displacement parameters are given by
\begin{eqnarray}
r_{1}=\frac{1}{4}\ln\left(\frac{4g_{2}}{\omega_{M}}+1\right),\hspace{1 cm}
\alpha_{1}=-\frac{g_{1}e^{-3r_{1}}}{\omega_{M}}.
\end{eqnarray}

\begin{figure}[tbp]
\center
\includegraphics[bb=30 20 474 322, width=0.47 \textwidth]{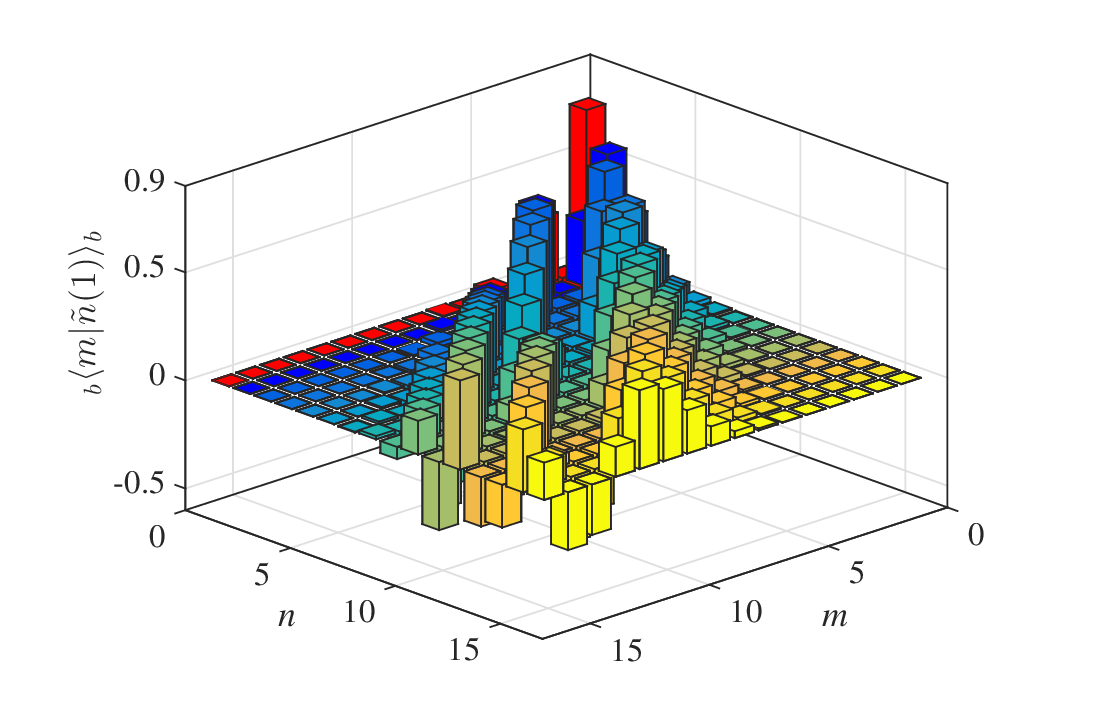}
\caption{The value of the overlap $_{b}\langle m|\tilde{n}(1)\rangle_{b}$ between the number state $|m\rangle_{b}$ and the single-photon squeezed displaced number state $|\tilde{n}(1)\rangle_{b}$ versus the state indexes $m$ and $n$. The used parameters are $g_{1}/\omega_{M}=0.8$ and $g_{2}/\omega_{M}=0.1$.}
\label{innerprod}
\end{figure}

The value of the inner product in Eq.~(\ref{innerprodxx}) can be calculated with the formula~\cite{Kral1990JMO}
\begin{eqnarray}
&&_{b}\langle m|\hat{S}(\zeta)\hat{D}(\beta)|n\rangle_{b}\nonumber\\
&=&\frac{1}{(m!n!\mu) ^{1/2}}\left(\frac{\nu }{2\mu}\right)^{m/2}\exp\left(-\frac{\vert\beta\vert ^{2}}{2}+\frac{\nu^{\ast}}{2\mu}\beta^{2}\right)\nonumber\\
&&\times \sum_{k=0}^{\min(n,m)}\frac{C_{n}^{k}2^{k}m!}{(m-k)!}
(2\mu\nu) ^{-k/2}H_{m-k}\left(\frac{\beta}{\sqrt{2\mu\nu}}\right)\nonumber\\
&&\times\left(-\frac{\nu^{\ast}}{2\mu}\right)^{(n-k)/2}H_{n-k}\left(\frac{\beta \nu^{\ast}-\beta^{\ast}\mu}{\sqrt{-2\mu\nu^{\ast}}}\right),
\end{eqnarray}
where $\zeta=se^{i\theta}$, $\mu=\cosh s$, and $\nu=\exp(-i\theta)\sinh s$. In this system, the transition matrix elements are determined by the overlap $_{b}\langle m|\tilde{n}(1)\rangle_{b}$ between the number state $|m\rangle_{b}$ and the single-photon squeezed displaced number state $|\tilde{n}(1)\rangle_{b}$. In Fig.~\ref{innerprod}, we plot the value of the overlap $_{b}\langle m|\tilde{n}(1)\rangle_{b}$ as a function of the two state indexes $m$ and $n$. Here we can see that for a given $n$, the dominate transition channel index $m$ is around the $n$. The value of the overlap could be either positive or negative, and the absolute value of the overlap decreases with the increase of the index difference $|m-n|$.

The single-photon emission spectrum can be calculated by evaluating the single-photon probability distribution in the continuous fields as $S(\Delta _{k})=$Tr$[|1_{k}\rangle _{c}\,_{c}\langle 1_{k}|\rho(\infty)]$~\cite{Liao2014PRA}, where $\rho(\infty)$ is the density matrix of the system in the long-time limit. When the mechanical resonator is initially in the pure states $\sum_{m_{0}=0}^{\infty}C_{m_{0}}(0) |1\rangle_{a}|m_{0}\rangle_{b}|\emptyset\rangle_{c}$ and the mixed states $\sum_{m_{0}=0}^{\infty}P_{m_{0}}(0)|m_{0}\rangle
_{b}\,_{b}\langle m_{0}|$, the single-photon emission spectrum can be calculated, respectively, with the following relations
\begin{subequations}
\label{spectrumform}
\begin{align}
S(\Delta_{k})&=\rho(\omega_{k})\sum_{m=0}^{\infty}\left\vert\sum_{m_{0}=0}^{\infty}C_{m_{0}}(0)B_{m_{0},m,k}\right\vert^{2},\\
S(\Delta_{k})&=\rho(\omega_{k})\sum_{m_{0}=0}^{\infty}P_{m_{0}}(0)\sum_{m=0}^{\infty}\vert B_{m_{0},m,k}\vert ^{2}.
\end{align}
\end{subequations}

\begin{figure}[tbp]
\center
\includegraphics[bb=13 118 283 475, width=0.47 \textwidth]{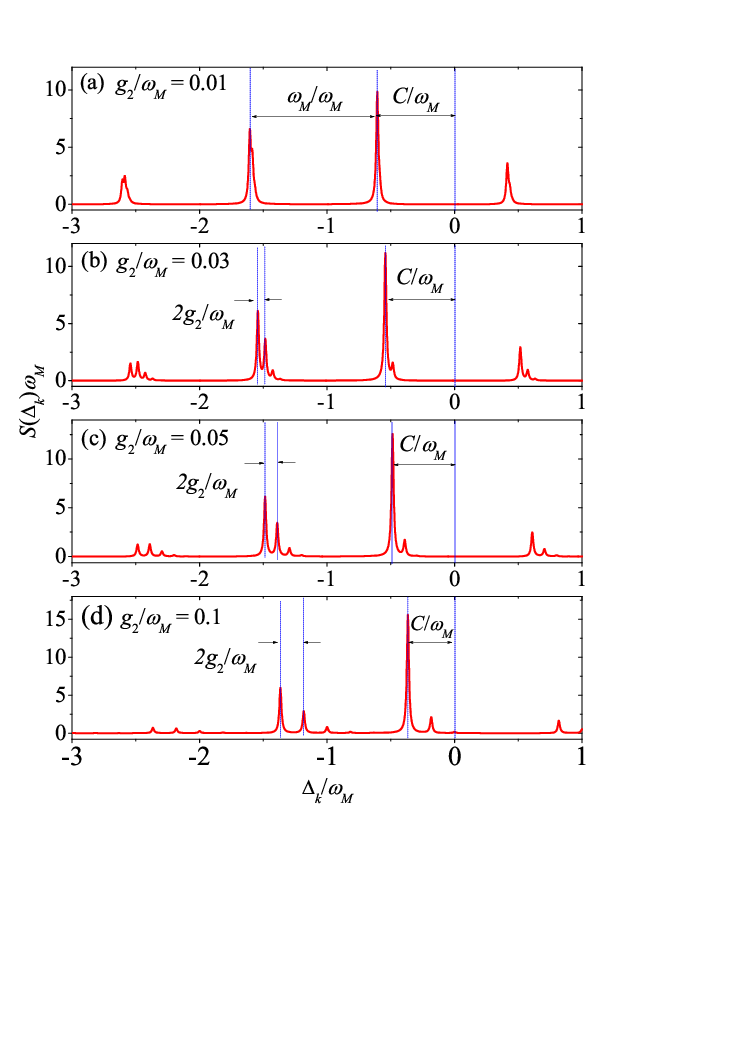}
\caption{(Color online) The emission spectrum $S(\Delta_{k})\omega_{M}$ as a function of $\Delta_{k}/\omega_{M}$ when the optomechanical-coupling strengths take different values $g_{2}/\omega_{M}=0.01$, $0.03$, $0.05$, and $0.1$. The initial state of the mechanical resonator is $|0\rangle_{b}$. Other parameters are given by $g_{1}/\omega_{M}=0.8$ and $\kappa/\omega_{M}=0.02$.}
\label{emicarig2}
\end{figure}

To observe the features of the single-photon spectra in this mixed optomechanical system, in Fig.~\ref{emicarig2} we plot the emission spectrum as a function of the photon frequency detuning $\Delta_{k}$ for various values of the coupling strength $g_{2}$ when the mechanical mode is initially prepared in its ground state $|0\rangle_{b}$. Similar to the radiation-pressure-type optomechanical coupling case, we choose proper parameters such that the system works in the single-photon strong-coupling regime $g_{1}>\kappa$ and the resolved-sideband regime $\omega_{M}>\kappa$. Then the spectral peaks corresponding to phonon sideband resonance become visible in the spectrum. Figure~\ref{emicarig2} shows that the quadratic optomechanical interaction will lead to the shift of the sideband peaks in the spectra. In particular, for a relatively large $g_{2}$, the sideband peaks will be splitted to many sub peaks. The appearance of these sub peaks is caused by the change of the mechanical resonance frequency in the presence of the quadratic optomechanical coupling. When there are $n$ photons in the cavity, the resonance frequency of the mechanical resonator is $\omega_{M}e^{2r_{n}}$. For the single-photon case, we have $\omega_{M}e^{2r_{1}}=\omega_{M}\sqrt{1+4g_{2}/\omega_{M}}$. Due to the difference between $\omega_{M}e^{2r_{1}}$ and $\omega_{M}$, there exist some sub peaks around the main phonon-sideband peaks, and the distance between two neighboring sub peaks is given by $\omega_{M}(\sqrt{1+4g_{2}/\omega_{M}}-1)$. The location of these phonon sideband peaks and sub peaks can be determined exactly by analyzing the resonance condition in Eq.~(\ref{resoncondi}).
As shown in Eq.~(\ref{resoncondi}), the peaks of these sidebands are located at $\Delta_{k}=E_{1,n}^{\prime}-E_{0,m}^{\prime}$, which leads to
\begin{eqnarray}
\Delta_{k}=\omega_{M}e^{2r_{1}}n-\omega_{M}m-C,
\end{eqnarray}
with
\begin{eqnarray}
C=g^{2}_{1}e^{-4r_{1}}/\omega_{M}-g_{2}e^{-2r_{1}}-\omega_{M}\sinh^{2}r_{1}\label{defofC}
\end{eqnarray}
being a constant energy shift of the ground state of the oscillator induced by the single-photon squeezing and displacement.
This energy shift $C$ can be read out from the spectrum, as shown in Fig.~\ref{emicarig2}(a). Here the first peak (starting at $\Delta_{k}=0$) in the red-sideband region corresponds to the zero-phonon line~\cite{Rabl2011PRL}, with the transitions from state $|1\rangle_{a}|\tilde{0}(1)\rangle_{b}$ to state $|0\rangle_{a}|0\rangle_{b}$.

When a single photon is placed inside the cavity, it will squeeze and displace the mechanical mode, and then the resonance frequency of the mechanical resonator will be changed to $\omega_{M}e^{2r_{1}}$. The phonon sideband peaks are determined by the transitions with different values of $m$ under a given $n$. Consequently, the distance between two neighboring main sideband peaks is $\omega_{M}$. In addition, the sub peaks are induced by these transitions relating to $m=n$. Therefore, the distance between the locations of two neighboring sub peaks is $\omega_{M}(e^{2r_{1}}-1)$. In the limitation of $g_{2}/\omega_{M}\ll1$, we have
\begin{equation}
\omega_{M}(e^{2r_{1}}-1)\approx2g_{2}.\label{2g2relation}
\end{equation}
It follows from Eq.~(\ref{2g2relation}) that an estimation of the coupling strength $g_{2}$ of the quadratic optomechanical interaction can be realized by analyzing the distance between two neighboring sub peaks. To resolve these sub peaks in the single-photon spectrum, the parameter condition $2g_{2}>\kappa$ should be satisfied.

Based on the discussions in the above sections, we can see that the energy shift $C$ and the coupling strength $g_{2}$ can be read out from the spectrum. According to the relation in Eq.~(\ref{defofC}) and the value of $g_{2}$, the value of the linear optomechanical coupling strength $g_{1}$ can be obtained. Therefore, the optomechanical coupling strengths $g_{1}$ and $g_{2}$ can be can be inferred with the spectrometric method. It should be pointed out that in our simulations we use the parameters corresponding to the single-photon strong-coupling regime, which is not accessible by current experimental conditions. Currently, the strengths of the optomechanical couplings are smaller than the decay rate of the cavity field in typical optomechanical systems~\cite{Aspelmeyer2014RMP}. Here, the single-photon strong optomechanical couplings are considered so that the phonon-sideband peaks and sub peaks can be resolved in the spectra, and then the results can show the physical mechanism and picture regarding the relation between the spectral feature and the optomechanical couplings.
\begin{figure}[tbp]
\center
\includegraphics[bb=11 118 282 474, width=0.47 \textwidth]{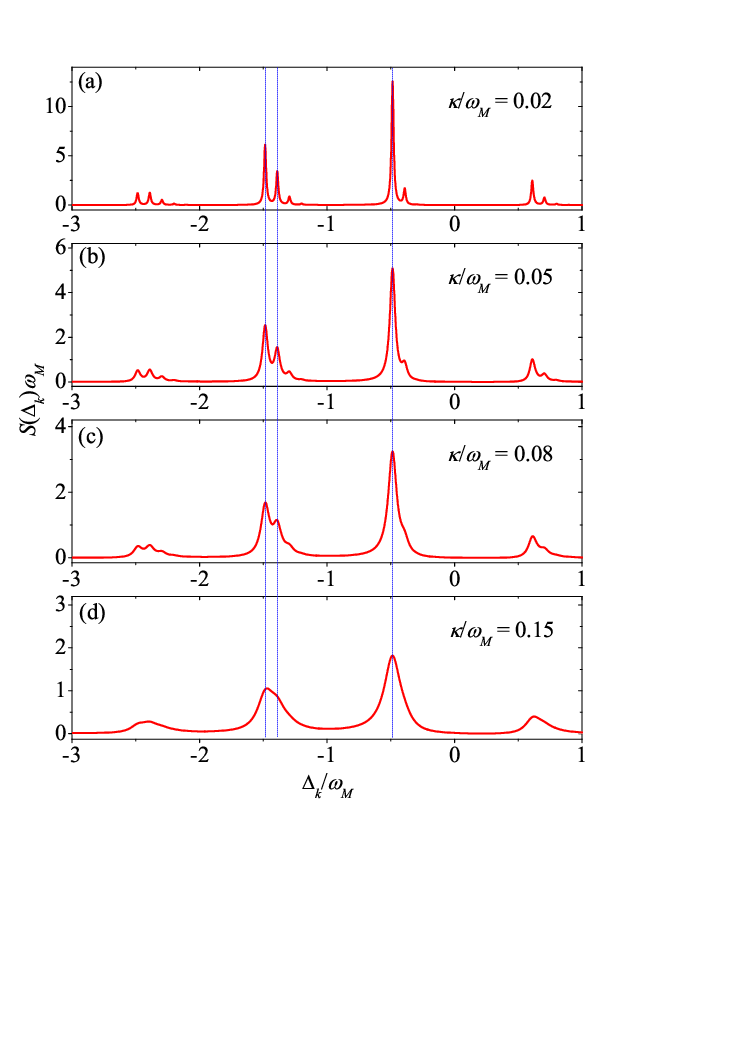}
\caption{(Color online) The emission spectrum $S(\Delta_{k})\omega_{M}$ as a function of $\Delta_{k}/\omega_{M}$ when the cavity field decay rate $\kappa$ take different values $\kappa/\omega_{M}=0.02$, $0.05$, $0.08$, and $0.15$. The initial state of the mechanical resonator is $|0\rangle_{b}$. Other parameters are given by $g_{1}/\omega_{M}=0.8$ and $g_{2}/\omega_{M}=0.05$.}
\label{emivarikappa}
\end{figure}

It has been shown that, to resolve these phonon-sideband resonance peaks in the single-photon emission and scattering spectra, the optomechanical system should work in the resolved-sideband regime $\omega_{M}>\kappa$~\cite{Liao2012PRA}. The parameter condition for resolving these sub peaks can be further confirmed by comparing the spectrum in different cases with various values of the cavity-field decay rates. In Fig.~\ref{emivarikappa}, we plot the single-photon emission spectrum as a function of the detuning $\Delta_{k}$ when the decay rate $\kappa$ takes different values: $\kappa/\omega_{M}=0.02$, $0.05$, $0.08$, and $0.15$. Here we can see that, with the increase of the cavity-field decay rate, the width of the sub peaks increases and then these sub peaks become indistinguishable. Physically, to resolve the sub peaks in these phonon sidebands, the decay rate should be smaller than the distance between two neighboring sub peaks. The corresponding parameter condition can be expressed as $\omega_{M}(e^{2r_{1}}-1)>\kappa$. For the parameters used in Fig.~\ref{emivarikappa}, the value of the distance between two neighboring sub peaks is $\omega_{M}(e^{2r_{1}}-1)\approx2g_{2}=0.1$. Then the sub-peak resolution condition is satisfied in Figs.~\ref{emivarikappa}(a-c). Accordingly, these sub peaks can be resolved in Figs.~\ref{emivarikappa}(a-c).

\begin{figure}[tbp]
\center
\includegraphics[bb=13 117 283 475, width=0.47 \textwidth]{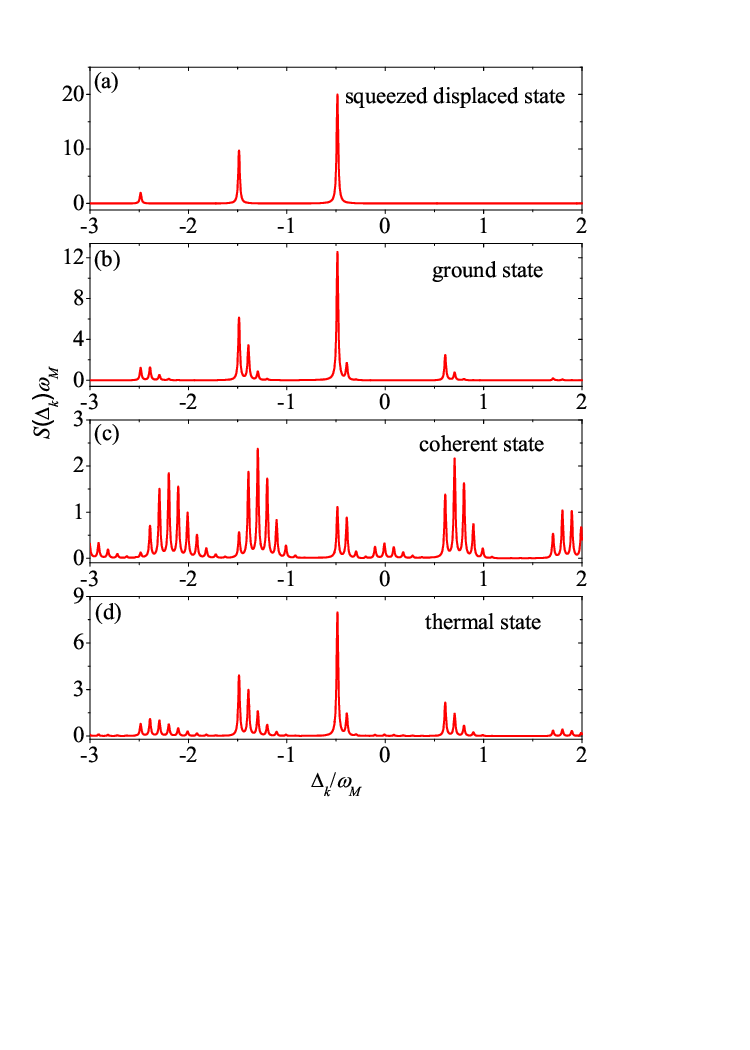}
\caption{(Color online) The emission spectrum $S(\Delta_{k})\omega_{M}$ as a function of $\Delta_{k}/\omega_{M}$ when the mechanical resonator is initially in different states: (a) single-photon squeezed displaced ground state $|\tilde{0}(1)\rangle_{b}$, (b) ground state $|0\rangle_{b}$, (c) coherent state $|\alpha\rangle_{b}$ with $\alpha=1$, and (d) thermal state $\rho_{\text{th}}$ with the average thermal phonon number $\bar{n}_{\text{th}}=1$. Other parameters are given by $g_{1}/\omega_{M}=0.8$, $g_{2}/\omega_{M}=0.05$, and $\kappa/\omega_{M}=0.02$.}
\label{inistatdepent}
\end{figure}
Physically, the locations of the phonon sideband peaks and the sub peaks are determined by the energy spectrum of the system.
However, the magnitude distribution of these peaks will depend on the initial state of the mechanical resonator. Below, we study how the emission spectrum $S(\Delta_{k})$ depends on the initial state of the mechanical mode. We plot in Fig.~\ref{inistatdepent} the emission spectrum $S(\Delta_{k})$ as a function of the detuning $\Delta_{k}$. We consider four different initial states of the mechanical mode: single-photon squeezed displaced ground state $|\tilde{0}(1)\rangle_{b}$ (i.e., the ground state in the squeezed displaced representation), ground state $|0\rangle_{b}$, coherent state $|\beta\rangle_{b}=e^{-|\beta|^{2}/2}\sum_{n=0}^{\infty}\frac{\beta^{n}}{\sqrt{n!}}|n\rangle_{b}$, and thermal
state $\rho^{th}_{b}=\sum_{n=0}^{\infty}\frac{\bar{n}_{\text{th}}^{n}}{(\bar{n}_{\text{th}}+1)^{(n+1)}}|n\rangle_{bb}\langle n|$ with $\bar{n}_{\text{th}}$ being the average thermal phonon occupation number. In Fig.~\ref{inistatdepent}(a), we can see that there are some peaks in the red sideband region ($\Delta_{k}<0$), and the distance between two neighboring phonon sideband peaks is $\omega_{M}$. Since the state of the mechanical resonator is $|\tilde{0}(1)\rangle_{b}$, then the transitions associated with the single-photon emission are from states $|1\rangle_{a}|\tilde{0}(1)\rangle_{b}$ to states $|0\rangle_{a}|m\rangle_{b}$, and there are no sub peaks in the spectrum. Differently, in Figs.~\ref{inistatdepent}(b-d), we can see that there exist phonon sideband peaks in both the red and blue sideband regions. This is because the mechanical resonator in the single-photon squeezed-displaced-number-state representation has some excited-state populations, as shown in the expansion $|n_{0}\rangle_{b}=\sum_{n=0}^{\infty}(\langle\tilde{n}(1)|_{b}|n_{0}\rangle_{b})|\tilde{n}(1)\rangle_{b}$. As a result, there will have some probabilities for the single photon absorbing the phonon's energy and leaving the cavity with a frequency larger than $\omega_{c}$, i.e., there are some peaks in the blue sideband region. Meanwhile, there are some sub peaks around the phonon sideband peaks. The number of these phonon sideband peaks and sub peaks depends on the initially contributing phonon distribution in the mechanical resonator. More peaks will be seen for a wider contributing phonon distribution in the initial state of the mechanical mode.

\section{Single-photon scattering\label{SPscattering}}

In the single-photon scattering case, we assume that the single photon is initially in a Lorentzian wavepacket outside the cavity, then the initial state of the system reads
\begin{equation}
|\varphi(0)\rangle=\sum_{k}\sqrt{\frac{\epsilon}{\pi\rho(\omega_{k})}}
\frac{1}{\Delta_{k}-\Delta_{0}+i\epsilon}|0\rangle_{a}|m_{0}\rangle_{b}|1_{k}\rangle_{c},
\end{equation}
where $\Delta_{0}$ and $\epsilon$ are the center and spectral width of the single-photon injection wavepacket, respectively. The corresponding initial condition reads $A_{m}(0)=0$ and
\begin{eqnarray}
B_{m,k}(0)=\sqrt{\frac{\epsilon}{\pi\rho(\omega_{k})}}\frac{\delta_{m,m_{0}}}{\Delta_{k}-\Delta_{0}+i\epsilon}.
\end{eqnarray}
Under this initial condition, the long-time ($\max\{1/\kappa,1/2\epsilon\}\ll t\ll1/\kappa$) scattering solution can be obtained as
$A_{m_{0},m}=0$ and (up to a phase factor $e^{-i(E_{0,m}^{\prime}+\Delta_{k})t}$)
\begin{eqnarray}
B_{m_{0},m,k}&=&\sqrt{\frac{\epsilon}{\pi \rho(\omega_{k})}}\left[\frac{1}{\Delta_{k}-\Delta_{0}+i\epsilon}\delta_{m,m_{0}}\notag\right. \\
&&\left.-i\kappa\sum_{n=0}^{\infty}\frac{1}{\Delta_{k}+E_{0,m}^{\prime}-E_{1,n}^{\prime}+i\kappa/2}\notag\right. \\
&&\left.\times \frac{\,_{b}\langle m|\tilde{n}(1)\rangle_{b}\,_{b}\langle \tilde{n}(1)|m_{0}\rangle_{b}}{\Delta_{k}-\Delta
_{0}+E_{0,m}^{\prime}-E_{0,m_{0}}^{\prime}+i\epsilon}\right].\label{Bmmkscat}
\end{eqnarray}
Similar to the emission case, the subscript $m_{0}$ in $A_{m_{0},m}$ and
$B_{m_{0},m,k}$ is used to mark the initial state $|m_{0}\rangle_{b}$ of the mechanical mode.
According to the probability amplitude~(\ref{Bmmkscat}) and the spectrum formula~(\ref{spectrumform}), we can obtain the single photon scattering spectrum. In Eq.~(\ref{Bmmkscat}), the first term  corresponds to the process in which the photon is reflected by the left cavity mirror without entering the cavity. The second term comes from the interaction process after the single photon entering the cavity. Note that the summation part in the second term of Eq.~(\ref{Bmmkscat}) has the same form as that appearing in single-photon emission process discussed above.
\begin{figure}[tbp]
\center
\includegraphics[bb=9 119 283 475, width=0.47 \textwidth]{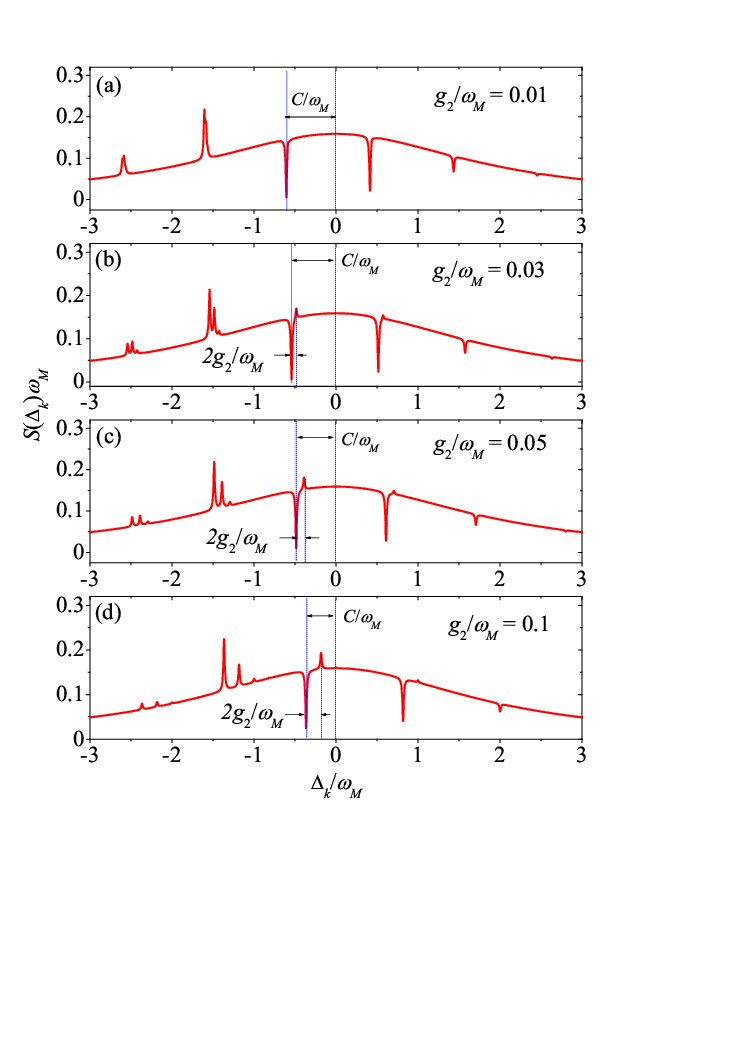}
\caption{(Color online) The scattering spectrum $S(\Delta_{k})\omega_{M}$ as a function of $\Delta_{k}/\omega_{M}$ in the wide-wavepacket-injection case when the optomechanical-coupling strengths take different values $g_{2}/\omega_{M}=0.01$, $0.03$, $0.05$, and $0.1$. The initial state of the mechanical resonator is $|0\rangle_{b}$. Other parameters are given by $\epsilon/\omega_{M}=2$, $g_{1}/\omega_{M}=0.8$, and $\kappa/\omega_{M}=0.02$.}
\label{scatwide}
\end{figure}

For the single-photon scattering case, the width of the injected single-photon wavepacket is a controllable variable. Here we focus on the wide wavepacket injection case $\epsilon/\omega_{M}>1$.
In Fig.~\ref{scatwide}, we plot the single-photon scattering spectrum $S(\Delta_{k})$ versus the photon detuning $\Delta_{k}$ for
various values of the coupling strength $g_{2}$ when the mechanical resonator is initially prepared in its ground state $|0\rangle_{b}$. Similar to the emission case, the phonon sideband peaks are visible when the system works in the single-photon strong-coupling and resolved-sideband regimes. In particular, the sub peaks can also be seen in the scattering spectrum. Different from the emission case, the spectrum show both peaks and dips in the scattering case. Physically, these tips are caused by quantum interference between the direct photon reflection channel and the scattering channel. The location of the first (starting from $\Delta_{k}=0$) dip in the red-sideband region and the distance between these sub peaks in Fig.~\ref{scatwide} can also be used to infer the coupling strengths $g_{1}$ and $g_{2}$. We also did the simulation for the narrow wavepacket injection case, and find that these sub peaks cannot be seen in the scattering spectrum. This is because the resonant narrow-wavepacket injection induce definite transitions, which dominates the single-photon transition process and then other transitions become weak in this system.

\section{Conclusion \label{Conclusion}}

In conclusion, we have studied analytically the single-photon emission and scattering spectra in a mixed optomechanical system. Based on the exact single-photon solutions, we have found the connection between the spectral
features and the optomechanical interactions. These spectra can also provide the signature of quantum optomechanical interaction between the photons and the mechanical oscillation at the single-photon level. By analyzing the phonon sideband peaks and sub peaks in the single-photon emission and scattering spectra, the coupling strengths of the optomechanical interactions can be inferred.

\begin{acknowledgments}
J.-F.H. is supported in part by the National Natural Science Foundation of China (Grant No.~11505055) and Scientific Research Fund of Hunan Provincial Education Department (Grant No. 18A007). J.-Q.L. is supported in part by National Natural Science Foundation of China (Grants No.~11822501 and No.~11774087), and Natural Science Foundation of Hunan Province, China (Grant No.~2017JJ1021).
\end{acknowledgments}

\end{document}